\documentclass{llncs}
\usepackage{makeidx}  
\usepackage[british]{babel}
\usepackage{url}
\usepackage[pdftex,colorlinks=true]{hyperref}
\usepackage{graphicx}

\title{Dear CAV, We Need to Talk About Reproducibility}

\author{Tom Crick\inst{1} \and Benjamin A. Hall\inst{2} \and Samin Ishtiaq\inst{3}}

\institute{Department of Computing \& Information Systems\\Cardiff Metropolitan University, UK\\
\email{tcrick@cardiffmet.ac.uk}
\and
MRC Cancer Unit, University of Cambridge, UK\\
\email{bh418@mrc-cu.cam.ac.uk}
\and
Microsoft Research Cambridge, UK\\
\email{samin.ishtiaq@microsoft.com}
}

\begin{document}
\frontmatter          
\pagestyle{headings}  

\maketitle

\begin{abstract}
How many times have you tried to re-implement a past CAV tool paper,
and failed?

Reliably reproducing published scientific discoveries has been
acknowledged as a barrier to scientific progress for some time but
there remains only a small subset of software available to support the
specific needs of the research community (i.e. beyond generic tools
such as source code repositories). In this paper we propose an
infrastructure for enabling reproducibility in our community, by
automating the build, unit testing and benchmarking of research
software.
\end{abstract}


\section{Introduction}\label{intro}


This is not a theory paper or a tool paper, nor an industrial case
study. The primary aim of this paper is to start a discussion in the
CAV community about the reproducibility of algorithms, models, tools
and benchmarks across the computer-aided formal analysis and
verification research domain. There is a significant opportunity for
the CAV community to identify and address the technical and
socio-cultural issues surrounding reproducibility in both the cognate
research domain as well as more broadly for computer science; a
desirable outcome would be a clear specification to encourage, enable
and enforce reproducibility. Enumerating a standard for
reproducibility would have a clear benefit for researchers as well as
the CAV community as a whole.

We recognise that we do not need to sell the idea of reproducibility
to the CAV community too much: we are used to writing papers about our
algorithms, models and tools; and reproducing others' work (or having
our own work reproduced) is encapsulated in the ``Benchmark Tables''
that CAV authors and referees both think are essential to any
paper. But the idea of reproducibility is gaining momentum in the
wider scientific community as well, for example in computer
science~\cite{collberg-et-al:2014}, life
sciences~\cite{rollins-et-al:2014},
psychology~\cite{chambers-et-al:2014} and the social
sciences~\cite{conte-et-al:2012}.  There has been a revolution in the
sharing and dissemination of published papers (\emph{open access}) and
the subsequent discussions relating to the sharing of protocols and
materials (\emph{open science})~\cite{rssaaoe:2012}. But the ability
of a researcher to take published results and data and reimplement the
described workflow remains
difficult~\cite{peng:2011,sandve-et-al:2013,wilson-et-al:2014}.  We
have previously documented some of the technical and cultural barriers
to reproducing work across computing and the computational sciences,
both in terms of the sharing of
algorithms~\cite{crick-et-al_wssspe2} and models and benchmark
sets~\cite{crick-et-al_recomp2014,crick+chuehong:2014}.


A number of high-profile computer science conferences, including PLDI,
POPL, SIGMOD, CGO, SPLASH and ECAI, now explicitly acknowledge the
importance of reproducibility (and repeatability, recomputability and
the multitude of `Rs' that underpin e-research), as well as promoting
community-driven reviewing and validation~\cite{fursin+dubach:2014}.
For many --- the first time for CAV this year --- this takes the form
of the author providing \emph{artefacts} (an accessible tool for
reproducing results) to evaluate. Journals such as Nature, PLoS
Computational Biology and Bioinformatics explicitly require that
source code and data is made available online under some form of open
source license. While these initiative are great, they are often
optional, seem piecemeal, and do little to enable verification or
validation of scientific results at a later stage. Even within the
same field, there are different ideas of what defines reproducibility.

This paper is thus a ``Call to Action'', inviting CAV practitioners to
embrace a new methodology for disseminating research. Practically, we
propose an initial specification for a reproducibility service for CAV. We
present the requirements of the prototype, and a suggested plan for
introducing the tool to the community. Finally, we highlight key
implementation issues relating to security and general applicability
which will need mitigating or resolving before widespread acceptance
by the research community.  The benefits for reproducibility testing
for a service are clear. In a sense, reproducibility here is an
extension of the standard practice of \emph{testing}.  Three features
distinguish our aims for reproducibility here: compilation and testing
in a new machine, a continuous integration strategy for code commits,
and the ability to add and remove benchmarks to the test set.

We have a running example based upon the BioModelAnalyzer
(BMA)\footnote{\url{http://biomodelanalyzer.research.microsoft.com/}}
tool, which has been described by the authors over a series of VMCAI
and CAV
papers~\cite{cook-et-al:2011,benque-et-al:2012,cook-et-al:2014}.
BMA is a tool for the development and analysis (simulation, model
checking) of a specific class of formal models for biology. The tool
specifically allows users to test for model \emph{stability}; that
is, a bespoke algorithm proves that for all initial states a model
always ends in a single fixpoint. We have chosen this example due to
our familiarity with the tool, and to highlight historical examples
where a reproducibility service would have supported both
toolchain development and algorithm discovery.

\section{A specification for reproducible computational science}\label{spec}

A service for reproducibility is intended to play three important
roles. It should:

\begin{enumerate}
\item Demonstrate that a piece of code can be compiled, run
and behaves as described, without manual intervention from the
developer.
\item Store and link specific artefacts with their linked
publications or other publicly-accessible datasets.
\item Allow new benchmarks to be added, by users other than
the developer, to widen the testing and identify potential bugs.
\end{enumerate}

\subsection{{\em de Novo} build environments}

A service such as this must require minimal developer intervention.
This serves multiple purposes -- through automation for example, the
service can be enabled to compile new code and test new benchmarks
trivially. This also forces the developer to make publicly available
their local workarounds (i.e. hacks). As such, this requires the
developer to make the project dependencies clearly available, and
enables future changes in the dependencies (such as a library update)
to be tested automatically too.

Throughout the lifetime of BMA, development has been shared between a
number of developers, working on different aspects of the tool. Work
in algorithm development focuses on adding new features to a command
line tool with few dependencies, aiding rapid development. In
contrast, the graphical model construction and testing environment has
typically been done by a single or pair of individuals.  This
necessarily required a number of dependencies, reflecting the use of
Azure (Microsoft's cloud computing platform) and Silverlight (a
framework for rich Internet applications).

In an early stage of development it was found that only a single
machine was capable of deploying the web service. This arose as the
developer responsible for writing and deploying the user interface had
run a series of commands necessary to run the mixture of 64- and 32-
bit components on Azure. These commands needed only be run once, and
went undocumented, thus needing to be rediscovered later when other
team members attempted to deploy.  These problems would be identified
trivially through the proposed service; such undocumented commands
would lead to all tests failing until explicitly added to the build
process.

\subsection{There are two types of people}

The service must fit easily into the developers workflow. As
noted in Section~\ref{rollout} we expect that there will be some costs
to the users in terms of the time required to ensure that the code
compiles and runs on the service. To minimise this, the service needs
to connect to standard code repositories, automatically detecting and
responding to new versions of the code and updates to dependencies,
running tests for every new code commit.

\begin{figure}[!ht]
	\centering
	\includegraphics[width=\textwidth]{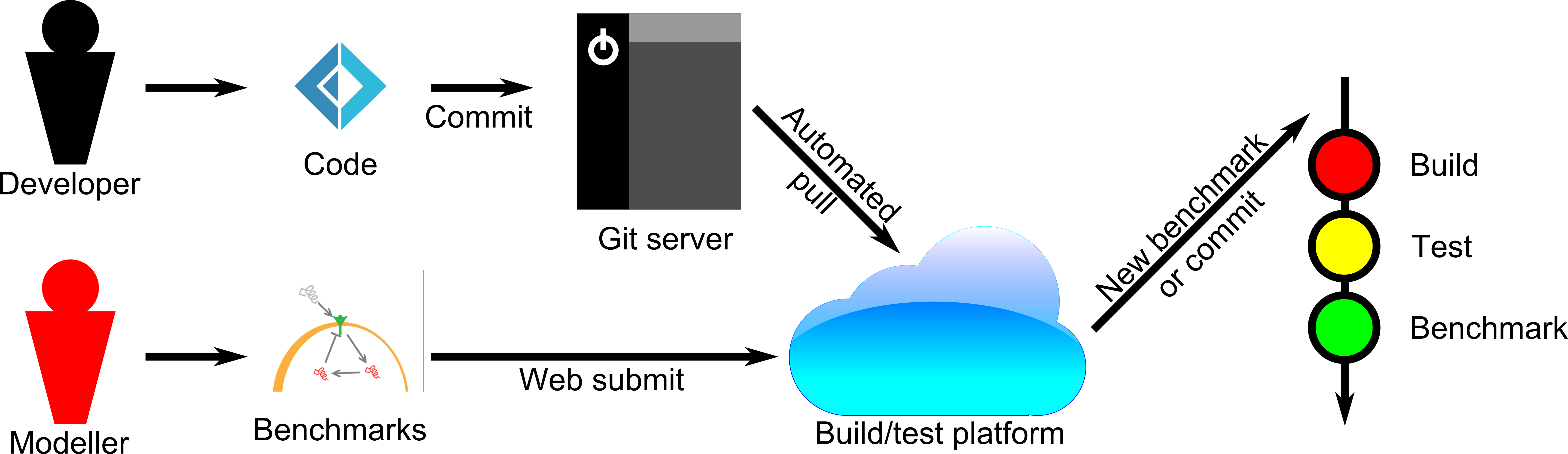}
	\caption{Proposed reproducibility service workflow}
	\label{schematic}
\end{figure}

To address these needs we propose that the service should follow the
workflow given in Figure~\ref{schematic}. Two classes of user are
defined; developers, who generate code; and modellers, who generate
new benchmarks. An individual might in practice play either or both of
these roles; here the roles serve to define ways in which people
interact with the system. A developer writes new code, which is
periodically pushed to a repository such as GitHub. Through
integration with the repository, the server responds to new code by
undergoing a process of pulling the code from the repository,
downloading required dependencies, and compiling the code. If this
stage fails, the developer is informed and the workflow ends. If the
code is successfully compiled, two stages of testing are
performed. The first stage (labelled {\emph{Test}} in
Figure~\ref{schematic}), involves running a series of basic tests
defined by the developer. This is intended as a sanity check to ensure
that basic features of the code have not been broken by the updated
code, and failure to pass these tests is reported and ends the
workflow. If this completes successfully, the second stage (labelled
{\emph{Benchmark}} in Figure~\ref{schematic}) a series of models are
tested for a known property, and the results recorded. These results
can then be stored in a database, with a note of the commit ID, and
available through a web interface for future analysis.

\subsection{Tool refinement}

In contrast to the developer role, a modeller supplies benchmarks for
a piece of code to test. These do not require that the latest version
of the code is recompiled, but on submission the models are tested and
added to the local repository of models for analysis.

In the case of BMA, throughout the development of the tool, many
refinements have been made to different implementations. Some of these
were subtle, and were identified by unit tests; for example rounding
mechanisms were switched between floors and rounds following a
scientific discussion. More complex changes however broke behaviours
which were not tested in our available benchmark set. One example was
in the treatments of nodes without inputs; ``biological intuition''
suggested that such nodes should have an alternative default function
from other nodes. Here, the ability of users to submit new benchmarks
would aid identification of these breaking changes, by extending the
test sets and simplifying the process of adding to the test sets, and
forcing the question of what changes are appropriate (and how to
update old models to keep correct behaviour).

\subsection{Identification of algorithmic weaknesses}

After a model is submitted, it is tested on every new piece of code
pushed to the server and the changes in the behaviour can be noted and
linked to specific code commits. Whilst the developer's role has a
transparent value (in providing an implementation of an algorithm),
the value of the modeller may be less immediately clear. The modeller
submits a broad range of tests which may highlight material flaws
(i.e. bugs) in the implementation, or the algorithm. More than this
however, the modeller may generate models which identify weaknesses of
either an algorithm or an implementation.

One example from the authors experience is the series of models with
``timed-switches'' described in~\cite{cook-et-al:2014}. There, we
presented a new algorithm for proving stability in a new class of
models. Whilst the paper focused on discussing the algorithm,
identifying the new class of models was complex. Models with long
cycles and the new class of models (``non-trivially stable models'')
both can take substantial time to search for cycles, and these models
could only be proved stable using a combination of simulation (to
identify the fix-point) and LTL queries (to prove that there existed
no paths beyond a certain length which did not include the
fix-point)~\cite{claessen-et-al:2013}.

\subsection{Algorithm--model axes}

The proposed service would allow both types of test to be included
explicitly, and models to be routinely added to each algorithm. Models
which time-out with one but are successfully proved can be logged and
identified for future study.  The features which define them would be
more easily found, and new algorithms developed to address the
specific features of the model. It could further be used to
demonstrate the speed improvement arising from new algorithms.

\subsection{{\texttt{makedepend}}}

Dependencies for a given implementation need explicit testing. Due to
the highly variable and sometimes complex nature of dependencies, we
see this as an optional part of the workflow, as developers may chose
to supply dependencies as binary files in the code compilation
process. For completeness however we note that such a system could
also respond to updates in external dependencies by triggering
compilation and testing in the same manner as defined for a new code
commit. This would aid developers in identifying code breaking changes
introduced by third parties.

\subsection{``I'm first!''}

Another issue is around performance comparisons of benchmarks: how can
we estimate and compare raw performance in the cloud? Testing new
algorithms on benchmarks is first about pass/fail, but very soon it's
about raw performance. Benchmark tables are about beating other
algorithms, other tools. But, if the whole verification workflow is
running on the cloud, then acquiring raw performance numbers is not
possible any more. There is no cloud equivalent of {\texttt{top}} or
{\texttt{time}} that gives user -- resource statistics. There is too
much infrastructure interference --- with VMs spinning up, being torn
down, migrating, the bus being used by other VMs, etc --- to obtain
faithful numbers for the user -- process -- VM itself. Projects such
as Recomputation.org\footnote{\url{http://recomputation.org/}}
have been focusing on using virtual machines in the cloud to freeze,
and later unfreeze, computational experiments; while this approach is
not the whole solution, it is certainly a move in the right
direction~\cite{arabas-et-al:2014}. Nevertheless, the project's
primary aim is to validate recomputation, with performance a secondary
consideration~\cite{gent:2013}. This is one aspect of this proposal
that needs further investigating.

\subsection{Running arbitrary code}

There are clearly security concerns around providing open
e-infrastructure that pulls, compiles and runs arbitrary
code as an autonomous continuous integration framework; we need to
consider precisely how this infrastructure would interact with other
open services, as well as privileges it would require to run as an
autonomous cloud service.

\section{A reproducibility model for the CAV community}\label{rollout}

Following the proposal of such a system, the question becomes:
{\emph{how do we encourage widespread uptake, or even standardisation?}}
Such a service may appear non-trivial, given the large numbers of
tools and workflows
that could potentially require to be supported by the service. Furthermore,
after such a service has been implemented, how do we ensure it is
\emph{useful} and \emph{usable} for researchers. To address this, we
propose the following workflow for CAV:

\begin{enumerate}
\item {\textbf{Pre-conference:}} clear signposting for authors; it
should be advertised and promoted in the CAV call for papers to
highlight this is a step-change in how we address
reproducibility. Call for artefact reviewers with a range of
specialisms, with a named chair of the review team.
\item {\textbf{Explicit criteria for authors:}} {\emph{make this as
easy as possible for us to evaluate/execute your artefact!}}. We would
aim to articulate the review criteria, but the primary aim is:
{\emph{can I evaluate/execute this artefact and get the same results
that are presented in the paper?}}
\item {\textbf{Submission:}} when papers are submitted, they have to
nominate whether they want their paper to go through artefact review
(at the start, this may not be compulsory, but this will change over a
period of time -- effecting cultural change and this would then become
a necessary condition and a formal stage in the reviewing workflow),
along with required tools, libraries and (ideally) computational
requirements.
\item {\textbf{Reviewing:}} in the first instance, it may be seen as
an extra (voluntary) step to the normal reviewing process:
e.g. {\emph{This submission is voluntary and will not influence the
final decision regarding the papers.}}. Independent of the scientific
merit of the paper, the results will be verified. To encourage this,
there may be a prize, as well as ranked ordering and profiled listed
in conference proceedings.
\item {\textbf{Artefact evaluation:}} artefact evaluation process runs
concurrent to the standard paper review process.
\item {\textbf{Reporting:}} traffic lights system (potentially with
ranked list) to indicate the level of reproducibility of the submitted
artefact.
\item {\textbf{Community curation:}} over a number of CAV cycles, we
would have a community curated repository/database of previous
artefacts, which would provide exemplars, comparisons and emerging
best practice.
\end{enumerate}

The key question for different research communities then becomes:
{\emph{how to initialise this change?}} Such a requirement creates a
set of new costs to researchers, both in terms of time spent ensuring
that their tools work on the centralised system (in addition to their
local implementation), but also potentially in terms of equipment (in
terms of running the system). Such costs may be easier to bear for
some groups compared to others, especially those with large research
groups who can more easily distribute the tasks, and it is important
that the service does not present a barrier to early career
researchers and those with efficient budgets (this type of cost
analysis is not unique to reproducibility efforts -- it has been
estimated that a shift to becoming exclusively open access for a
journal may lead to a ten-fold increase in computer science
publication costs~\cite{vardi-cacm-2014}).

\section{Conclusions}\label{concl}

The benefits to the community from a cultural change to favour
reproducibility are clear and as such we should aim through the
software infrastructure and the CAV workflow to mitigate these
costs. Furthermore, we can reasonably expect the needs of the
community to evolve over time, and initial implementations of the
platform may require refinement in response to user feedback. As such,
if the community is to move to requiring reproducibility, it seems
most reasonable that this is staggered over a number of years to allow
for both of these elements to develop, until eventually all authors
are required to use the service. This plan balances competing needs
within the community, and would reduce the disruption for uptake by
gradually introducing it to researchers.

\begin{description}
\item[CAV 2016:] Offer the service as an optional extra in the testing
phase, allowing users to demonstrate the reliability of their code
which could be taken into account in the review process.
\item[CAV 2017:] All authors must use the reproducibility service, but
results are not used in the review process. The results of the test
are used to refine the service and pick out any unaddressed issues
\item[CAV 2018:] All authors are required to use the service, and the
results are explicitly used to assess reproducibility in the review
process.
\end{description}

An open discussion and understanding of what reproducibility means for
the CAV community is important: we need to explicitly state that this
is worthwhile and address it, or don't bother doing it at all. We thus
propose that this should be trialled for CAV in 2016, with an
explicitly-defined three year schedule for universal adoption by CAV
2018.

\bibliographystyle{splncs}
\bibliography{cav2015}

\end{document}